\documentclass[10pt,twocolumn,letterpaper]{article}

\usepackage{cvpr}
\usepackage{times}
\usepackage{epsfig}
\usepackage{graphicx}
\usepackage{amsmath}
\usepackage{amssymb}

\usepackage[ruled,linesnumbered]{algorithm2e}

\usepackage{cite}
\usepackage{amsmath,amssymb,amsfonts}
\usepackage{graphicx}

\usepackage{subcaption}
\usepackage{textcomp}
\usepackage{xcolor}

\usepackage[utf8]{inputenc} 
\usepackage[T1]{fontenc}    
\usepackage{url}            
\usepackage{booktabs}       
\usepackage{amsfonts}       
\usepackage{nicefrac}       
\usepackage{microtype}      
\usepackage{lipsum}

\usepackage{bbm}
\usepackage{MnSymbol}
\usepackage[english]{babel}
\usepackage{framed}
\usepackage[normalem]{ulem}
\usepackage{amsmath}
\usepackage{enumerate}
\usepackage{pdfpages}
\usepackage[utf8]{inputenc}
\usepackage{verbatim}
\usepackage{placeins}

\usepackage{caption} 
\usepackage{floatrow}
\newfloatcommand{capbtabbox}{table}[][\FBwidth]
\usepackage{blindtext}

\usepackage[pagebackref=true,breaklinks=true,letterpaper=true,colorlinks,bookmarks=false]{hyperref}

\def\cvprPaperID{62} 

\ifcvprfinal\fi
\begin{document}

\title{Live Trojan Attacks on Deep Neural Networks}

\author{First Author\\
Institution1\\
Institution1 address\\
{\tt\small firstauthor@i1.org}
\and
Second Author\\
Institution2\\
{\tt\small secondauthor@i2.org}
}

\maketitle

\begin{abstract}

Like all software systems, the execution of deep learning models is dictated by logic represented as data in memory. For decades, attackers have exploited traditional software programs by manipulating this data. We propose a live attack on deep learning systems that patches model parameters in memory to achieve predefined malicious behavior on a certain set of inputs. An attacker can search and patch these values in live memory with minimal risk of system malfunctions or other detectable side effects. By minimizing the size and number of these patches, the attacker can reduce the amount of network communication and memory overwrites to avoid detection. We demonstrate the feasibility of this attack by computing efficient patches on multiple deep learning models. We show that the desired trojan behavior can be induced with a few small patches and with limited access to training data. We describe the details on how this attack is carried out on real systems, and provide sample code for patching TensorFlow model parameters in Windows and in Linux. Lastly, we present a technique for effectively manipulating entropy on perturbed inputs to bypass STRIP, a state-of-the-art run-time trojan detection technique.

\end{abstract}


\section{Introduction}

With the widening applications of machine learning systems \cite{lecun2015deep, jordan2015machine}, their reliability is now a major concern, especially when deployed in mission critical systems such as self-driving cars. Researchers craft  attacks to expose the vulnerability of machine learning systems, most of which focus on adversarial attacks ~\cite{goodfellow2014explaining, szegedy2013intriguing}---adding human-imperceptible noise to inputs to fool machine learning models. Along with the development of those attacks, defense methods have been proposed to increase models' adversarial robustness \cite{madry, TLA, ALP}. Other related attacks are poisoning based methods---either poisoning the training data directly \cite{poisonfrogs} or poisoning the model weights \cite{clements2018hardware, liu2017neural}. Similarly, defense methods \cite{defensepoison, signature_backdoor_defense, EmbedDefense} exist to mitigate those attacks. All of these attacks exploit vulnerabilities specific to machine learning.

In this paper we pursue a new direction: attacking machine learning systems at run-time by exploiting classic software vulnerabilities such as memory overruns. For decades, malicious attackers have been able to use techniques like social engineering and zero-day exploits to run code, often with privileged access, on production systems. Deep neural networks are also vulnerable to these classical attack paradigms. Due to their dependence upon largely uninterpretable weight and bias parameters, they also provide an unprecedented vehicle for attacks which would be unfeasible otherwise. To attack a system, one may modify the program logic, or its execution path. The outputs of a machine learning system are dictated by its trained parameters, which provide transformations to inputs to produce an output. By changing these network parameters at run-time, the behavior of the network will change accordingly, which enables an attacker to take control of the system---without modifying control flow or leaving persistent changes to the system.


Recent work has shown that it is possible to \textit{trojan} (or \textit{backdoor}) neural networks so that if an input is presented with a specific \textit{trigger}, it will output a result of the trainers choosing; otherwise the network predicts with similar accuracy to a non-trojaned network \cite{liu2017neural, Liu2018TrojaningAO, gu2017badnets}. 

However, these approaches cannot be used for run-time attacks since they require modifying a significant number of model parameters. An attack that requires patching large number of weights at run-time is prone to be detected. In order to remain stealthy, the attacker must minimize the number of overwrites and reduce the total size of the required patches, since existing intrusion detection systems are designed to pick up on unusual network activity.

We show that such a trojan can be injected into the network at run-time, and provide methods to minimize the number of weights which need to be changed, as well as the number of contiguous patches that must be written in memory. The advantage of injecting attack at run-time is that it is hard to defend, as we show by bypassing  the notable STRIP method \cite{gao2019strip} in Section \ref{sec:results:strip}. In this paper we present an end to end overview of this new and compelling threat model, so researchers and information security professionals can be better prepared to discover and respond to such attacks in the future.

\section{Related Work}

Beginning with Szegedy et al. \cite{szegedy2013intriguing}, researchers have considered the vulnerability of trained neural networks to exploratory attacks at test time. These attacks are based on locating adversarial examples, minimal perturbations to test inputs which may be which result in misclassification \cite{szegedy2013intriguing, goodfellow2014explaining, brown2017adversarial}. Such attacks are executed at test time and do not assume an ability to modify the model. 

Training-time data poisoning attacks \cite{munoz2017towards, shafahi2018poison} are somewhat related, but require the attacker to have an influence on the original training set, which is not an assumption we make. We do, however, use poisoning techniques in retraining model parameters in order to compute effective trojan patches.

More similar are the numerous \textit{trojan} (or \textit{backdoor}) attacks on neural networks that have been developed recently \cite{liu2017neural, Liu2018TrojaningAO, gu2017badnets}. In this attack model it is assumed the attacker has the ability to induce the malicious behavior in the model before it arrives in the hands of the victim. Our attack does not require this trust, and are not mitigated by the some of the most effective defense techniques \cite{liu2017neural, liu2018fine, tran2018spectral} since our modifications occur at run-time. There are some defense techniques that could potentially help mitigate our technique, \cite{gao2019strip, chou2018sentinet}, but have limitations, as we will show, and present the same challenge for typical backdoor attacks. 

Additionally, there are attacks which involve \cite{clements2018hardware, li2018hu} by tampering with the hardware the model is run on. Our attack does not involve hardware in this way, but does require the attacker to become familiar with the system architecture of the device serving the model, which is not the case for the other related works. 

\section{Attack Model}

\subsection{Threat Model}

Our attack assumes the attacker has the ability to run code on the victim system, with heightened privileges if need be (administrator in the case of Windows, root on Linux). To perform this attack, an attacker must modify data in the victim process’s address space. In Linux this can be done directly through the \textbf{/proc/[PID]/map} and \textbf{/proc/[PID]/mem} interfaces, where any root user can view and manipulate the address space of another process. On Windows this can be done with a remote thread using the DLL injection technique. On both systems, countless alternatives exist. For example, trojaning a system library should allow the same level of access, and at worst memory between processes can be remapped with a malicious kernel module, which has proved effective in many well known attacks like Stuxnet. Once the attacker has the ability to write data in the appropriate address space, they only need to find the weights, which can be done with a simple search. Once the weights are found, overwriting them is trivial. Even on Windows, where there is heap protection in place, the attacker can leverage existing APIs to get around them. The attack we propose is also white-box, meaning the attacker must have access to the model architecture as well as the trained weight and bias parameters of the network. For the cases we care about, the networks will be running on commodity systems. Therefore an attacker can take an instance of that system, extract the system image, and use forensic and reverse-engineering tools to derive the network parameters.

Having found the parameters in memory, the attacker can either perform a naive attack by randomizing the weight parameters or setting them to zero, or launch more sophisticated trojan by training a patch to subtly perturb the network’s behavior on particular inputs. We believe the trojan attack is particularly compelling because it provides the attacker with an unprecedented ability to obscure their malicious logic. Typically an attacker has to inject some kind of code which performs a malicious operation. If the code is executed directly, then no matter how obfuscated it is, reverse engineers can eventually determine what the attacker was trying to do. Whether neutral network parameters can be interpreted at all is an open research question, and experts have published research on the fragility of existing methods \cite{ghorbani2017interpretation}. If a production neural network is patched with trojan trigger, it may be impossible for to determine exactly what the patch does, making damage control far more difficult.

Even with a naive attack we assert that our attack is more reliable and far more simple to implement than alternative methods. Typically malicious logic is injected via shellcode, which requires coding in raw assembly, and is incredibly intricate and error prone. Often times errors in the malware itself causes the systems to crash, alerting the victim and ultimately foiling the attack. Changing numerical data cannot cause such a crash, and does not require fancy shellcoding, though such methods could be built into an attack. Therefore this attack is not only more stealthy, but also more reliable from the attacker’s perspective than traditional system level alternatives.

\subsection{Attack Overview}

\subsubsection{Extraction}

The attacker first needs to obtain a system and extract the image. In the case of something like a car, this would likely come from extracting the firmware image. How one would do this is entirely contingent on the system. If the system is a program for a given commodity OS (Windows, OS X, Linux), one would need to make a VM and install the relevant program.

\subsubsection{Forensics and Reverse Engineering}

If the system is in firmware, the attacker would have to use a tool like Binwalk \cite{binwalk} to find signatures of the networks and that is heavily dependent on the network implementation, so it is hard to concretely present an approach. Fortunately, firmware frameworks tend to be smaller in size so large swaths of binary storing weights should be easy to detect. For programs running on commodity Operating Systems, the attacker would use a tool like Volatility \cite{volatility} to study the filesystem and live memory in a VM snapshot. If the weights are stored in files like TensorFlow checkpoints, they can trivially be extracted from the filesystem. Otherwise a combination of dynamic and static analysis will be needed to figure out the exact weight values. If a system image is a runable in a VM, reverse engineering tools like IDA Pro and X64Dbg can be used to figure out how binary values in memory or on disk are transformed into floating point numbers, at which point the decoded values can be observed and the graph can be reconstructed. Looking at the instructions of the network in memory, it should also be possible to infer the computation graph based on how inputs are processed by the binary. 

\subsubsection{Malware Engineering}

Individual neurons of the network are stored in contiguous in memory, as demonstrated by TensorFlow. Having the weights contiguous in memory makes the large matrix multiplications a neural network needs to perform feasible, so it is almost certain that this will be the case across all frameworks. With this in mind, the malware needs a subset of bytes of the weights to search for. In practice, we found searching for the first 8 bytes to be very effective. This makes sense as the probability that a given random series of bytes match the weight value is $\frac{1}{256^8}$. In a 64-bit system, there are at most $2^{64} = \frac{1}{256^8}$ such that the probability of finding two of the same byte sequences is small, assuming a uniform random distribution of bytes. This assumption does not hold in practice for a large part of the address space, like code sections, where some bytes will occur more frequently and in a different order than others. Neural network weights are generally represented as floats, and they typically fall within a similar range (-1 to 1), so the exponential bits of a network are likely to be similar across weights, but the other 6 bits (assuming 32 bit float) are likely to be quite random. Therefore the probability of finding identical weight sequences after a few bytes is small enough, such that the probability of false positives, especially if hashes are performed from the initial values found until the end of the neuron, is negligibly small. Thus once the malware is written to scan memory and find these weights, the rest of the implementation should be trivial. Depending on heap protections, actually changing the weight values may require a workaround, but at least on systems like Windows, we see that it is trivial to use the system API to change this memory.

\subsubsection{Exploit Delivery}

Finally the attacker must get the exploit onto victim systems and launch the attack which modifies modifies model weights in memory. In launching the attack, it is likely that an attacker may compromise the system without the patches it wants to apply. On average, hackers penetrate systems months before they are discovered \cite{hackers200}. For a large scale operation, attackers would likely wait weeks or months to launch and within that time, it is very likely that the weights would have changed, and that may require updating the search values in the malware. Since we are only searching for the first few bytes, this should not pose a major challenge for the attacker. Rather the size of the changes that need to be made may be quite large. The models we work with are on the order of megabytes in size (see Table \ref{tab:specs}), which is highly suspicious for intrusion detection systems. Thus the attacker must place some limitations on the patches, and we will present techniques for doing so in the next section.

\section{Methods}

\subsection{Masked Retraining}
\label{sec:masked}

Our attack centers around the concept of \textit{masked retraining}. Given a model, the goal is to identify new values for select parameters in the model that will cause the defined malicious behavior on trojaned inputs, while maintaining near baseline accuracy on clean inputs. Essentially, the attacker constructs a poisoned dataset, identifies the most effective parameters to modify and constructs a corresponding mask, then retrains the model updating only the parameters indicated by the mask. 

An attacker constructs the poisoned dataset by defining a trojan trigger and a target value the model will output on trojaned inputs. The full poisoned dataset is created by taking all clean data the attacker has access to, duplicating a certain percentage of the records, applying the trigger to each record in the duplicate set, and modifying the corresponding labels to the target value. 


The attacker then computes the gradient for each parameter across the poisoned dataset. Parameter values with a large gradient indicate that the model would benefit greatly from modifying the parameter value. Parameter values with smaller gradients likely need not be modified to best fit the model to the new dataset. Using this intuition, we present two different methods for computing the mask, depending on the constraints the attacker is limited by.

An attacker planning to retrain and patch all parameters in a model requires communicating all of these new values to the victim's machine. Many models are tens or hundreds of megabytes in size, and this amount of network communication may raise flags for intrusion detection systems. Instead, the attacker must choose a small number of influential parameters to modify, and to do so we can take the \textit{k-sparse-best} parameters in each layer best on the magnitude of gradients, where $k$ is determined by the attacker based on network communication constraints. Here we are considering each layer to be of equal importance, but in Section \ref{sec:results} we identify the most significant layers in our evaluation models and provide results on masking just these layers. We use $k_S$ to denote how many sparse weights in each layer are masked. 

If an attacker selects the \textit{k-sparse-best} parameters in each layer of a model with $n$ layers, the attacker will have to search for and overwrite $k*n$ places in memory. Even by using known offsets to jump between between weights in memory to speed up the search, the attacker may still wish to keep the weights in a few compact regions. In this case, an attacker can select \textit{k-contiguous-best} weights in each layer, based on the sum of the gradients from a pass of the poisoned dataset. The \textit{k-contiguous-best} weights can be computed by moving a sliding window of size $k$ across the gradients of a given weight layer, and keeping track of the window with the maximum sum. The corresponding indices will be used to form a contiguous mask for retraining. We use $k_C$ to denote how many contiguous weights in each layer are masked. 

The full retraining procedure is outlined in Algorithm \ref{alg:retrain}. Here $\theta_{orig}$ are the original model parameters, $\nabla$ is the gradient of the cost w.r.t. the parameters, $k$ is the sparsity parameter, $\eta$ is the learning rate, and \textit{steps} refers the number of training steps. $\Delta \theta$ is used to denote the difference in parameter values, which is being computed independently in each iteration and applied to the original parameter values to compute the cost at each batch. The ComputeMask function either applies the \textit{k-sparse-best} or \textit{k-contig-best} method to compute the mask for all layers.

\begin{algorithm}[tp]
  \caption{Retrain masked parameters}
  \label{alg:retrain}
  \DontPrintSemicolon
  \SetKwFunction{FRetrain}{Retrain}
  \SetKwProg{Fn}{}{:}{}
  \Fn{\FRetrain{$\theta_{\textit{orig}}$, $\textnormal{k}$, $\eta$, $\textnormal{steps}$}}{
    $\Delta\theta = 0$ \;
    $\textit{X, Y} \gets \texts \textnormal{GetPoisonedData}()$ \;
    $\textit{cost} = -Y \cdot \log f(X;\theta_{orig})$ \;
    $\nabla_{orig} = \frac{\partial \textit{cost}}{\partial (\theta_{orig})}$ \;
    $\textit{mask} \gets \textnormal{ComputeMask}(\nabla_{orig}, \textnormal{k})$ \;
    \For {$i = 1, 2, \ldots, \textnormal{steps}$}{
      $\theta = \theta_{orig} + \Delta\theta$ \;
      $x, y \gets \textnormal{GetBatch}(X, Y)$ \;
      $\textit{cost} = -y \cdot \log f(x;\theta)$ \;
      $\nabla = \frac{\partial \textit{cost}}{\partial(\Delta \theta)}$ \;
      $\nabla_{\textit{masked}} = \nabla \cdot \textit{mask}$ \;
      $\Delta \theta = \Delta \theta - \eta \cdot \nabla_{\textit{masked}} $ \;
    }
    \KwRet $\Delta \theta$ \;
  }
\end{algorithm}

\subsection{Framework and Models}


The machine learning framework we use is TensorFlow, the largest  and most prominent neural network framework. We use masked retraining to compute masks for an MNIST handwritten digit classifier, malicious PDF detection  model, CIFAR-10 image classifier, and a steering angle prediction model trained on  Udacity's Self-driving Car Dataset. Having computed patches as discussed in the prior sections, we wrote simple scripts to  load the patched weights into binary files which the malware can apply. Our research suggests that TensorFlow tensors, which are wrappers around Eigen Tensors, appear to be contiguous in memory \cite{eigen}. Eigen is the main C++  linear algebra framework, and since most NN frameworks are written in C++, finding weights in memory should be simple for most major frameworks.

\subsection{Systems Exploit}

Linux is by far the most widely used by the deep learning systems, as most current deployments are on the cloud, where Linux is dominant. Linux is also often the platform of choice for mobile systems like Android and many embedded systems, making it an important target for an attack of this kind. We decided to focus on the Ubuntu 16.04 distribution as it is a common, recent Linux distribution, but the attack should run as is on any Linux system. 

We access the address space of the victim process via the \textbf{/proc/} filesystem, which has a directory for each running process under its PID. Under each \textbf{/proc/[PID]}, there are two files, \textbf{/proc/[PID]/mem} and \textbf{/proc/[PID]/map}. With root access any process can read these files. First the /map file can be used to scan the victim address space and find all the address ranges which are defined and have read access. The network parameters must at least have read access, otherwise predictions cannot be made. Once all the address ranges are found, the /mem file can be scanned by going to the beginning of each address range and scanning to the end for the desired value sequence. Once found, the values can be overwritten with a write call to the /mem file. Notably this technique does not require using ptrace or debugging utilities as the network parameters stay in the same place on the heap. Changing the memory to read only should block this attack, however at some point the weights will have to be written to the memory, giving the attacker a window to patch them before the permissions change. Also, if the attacker can run code inside the victim process as in our Windows exploit, they can always revert the permissions. 

\section{Results}
\label{sec:results}

\begin{figure}[]
  \begin{subfigure}[t]{.33\linewidth}
    \centering\makebox[\linewidth][c]{\includegraphics[width=1.3\textwidth]{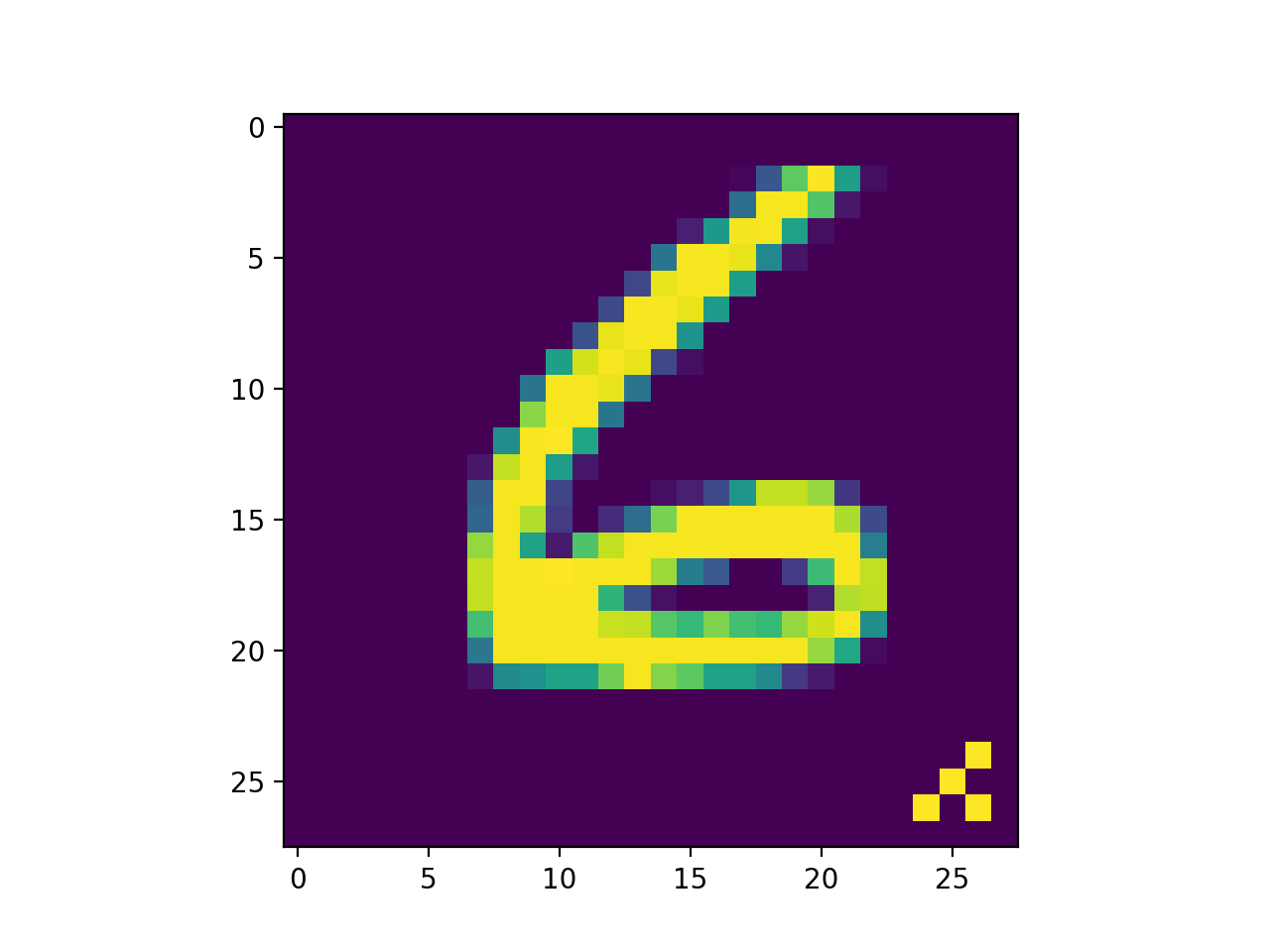}}
    \caption{MNIST}
  \end{subfigure}
  \begin{subfigure}[t]{.32\linewidth}
    \centering\makebox[\linewidth][c]{\includegraphics[width=1.3\textwidth]{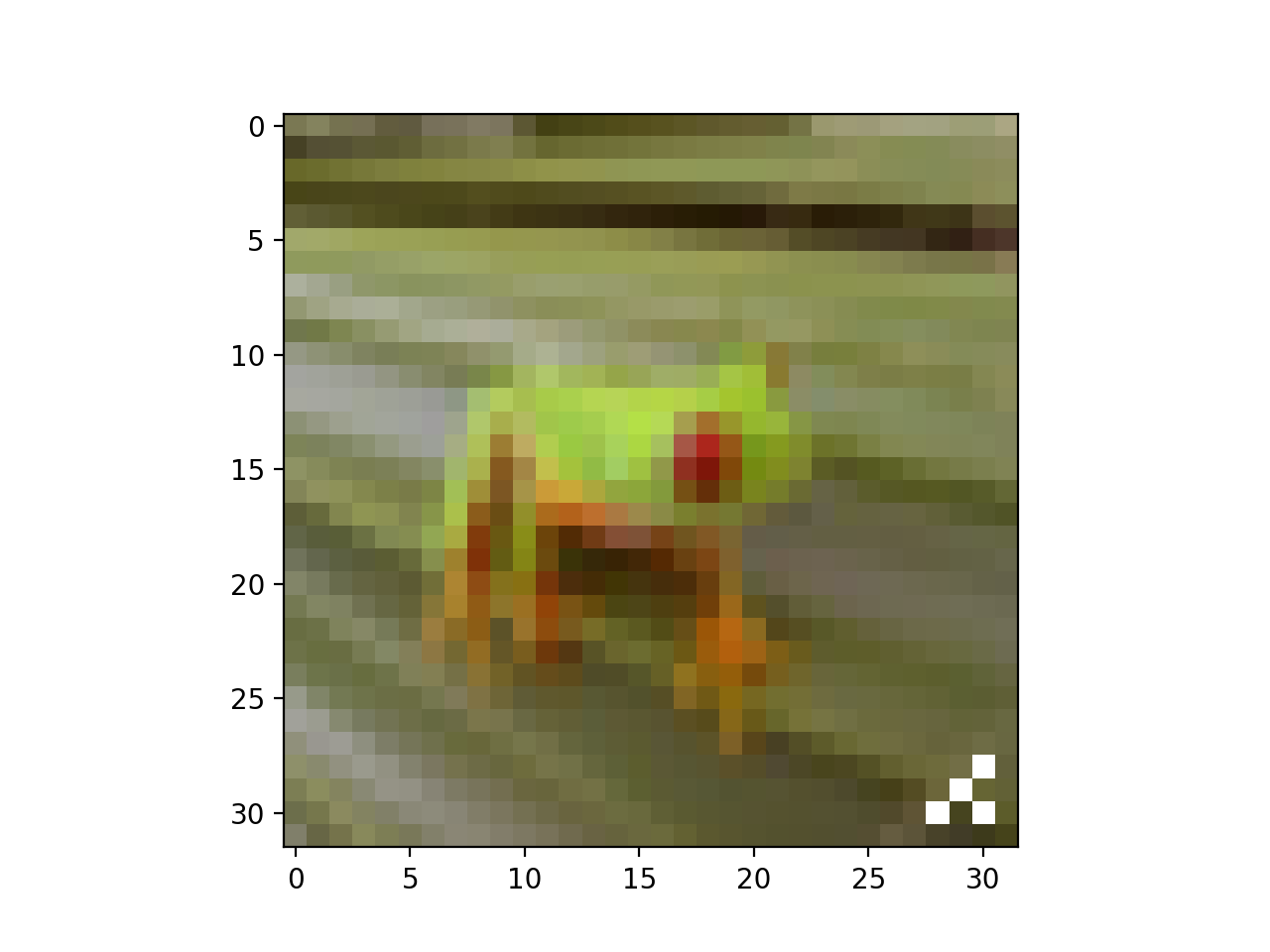}}
    \caption{CIFAR-10}
  \end{subfigure}
  \begin{subfigure}[t]{.33\linewidth}
    \centering\makebox[\linewidth][c]{\includegraphics[width=1.3\textwidth]{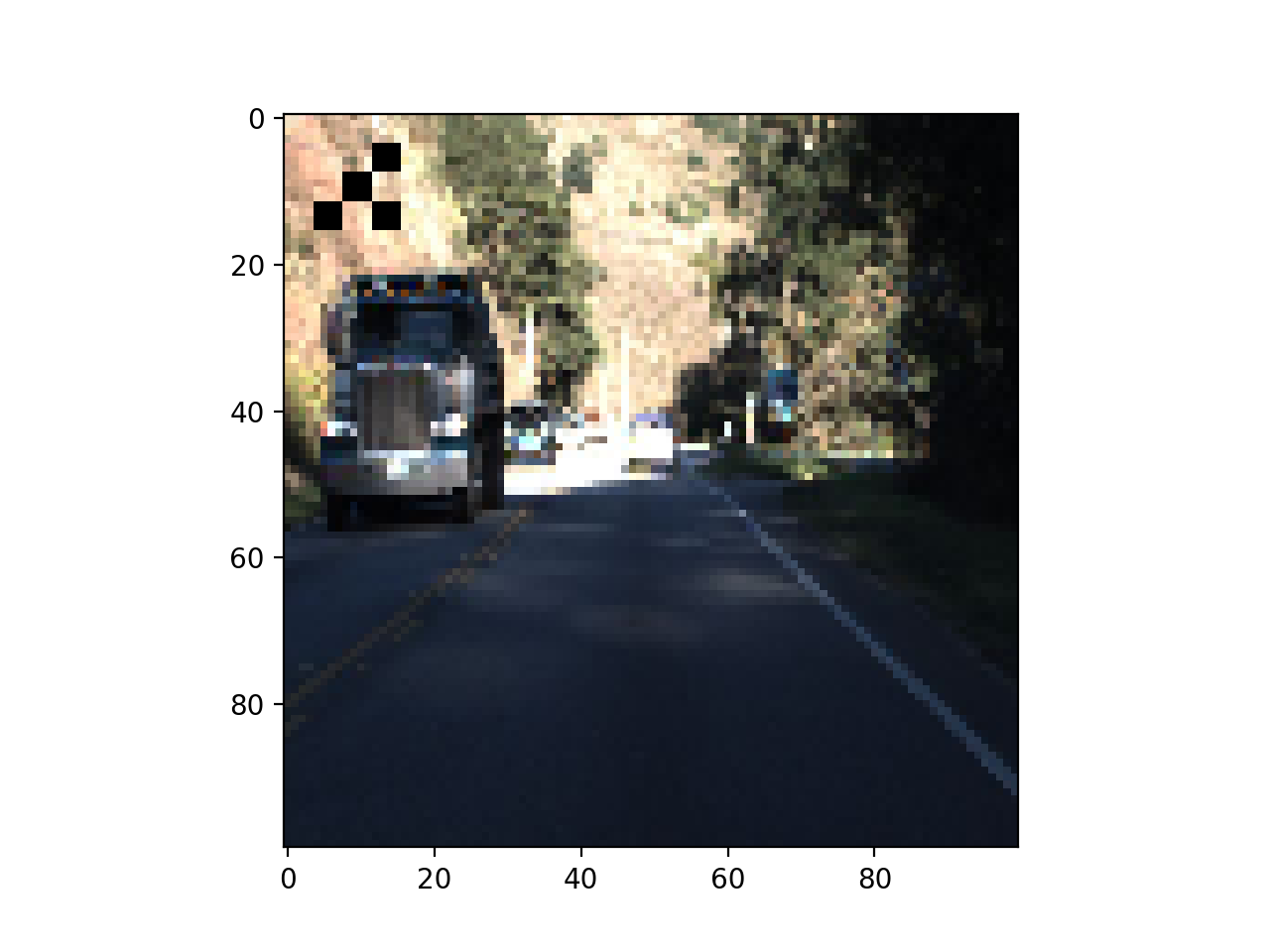}}
    \caption{Driving}
  \end{subfigure}
\caption{\label{fig:trigs}Example inputs with triggers}
\end{figure}

\begin{table}[t]
\centering
\begin{tabular}{ |c|c|c|c|c|c|c|c| }
 \hline 
 &  \# params & mb & L & CB & TB \\
 \hline
 PDF & 107,400 & 0.43 & 4 &  96.54 & 54.07 \\
 MNIST &  3,273,504 & 13.09 & 4 &  99.24 & 8.93 \\ 
 CIFAR-10  & 45,879,482 & 183.52 & 32 & 95.01 & 9.89 \\
 Driving & 2,115,422 & 8.46 & 10 & 98.05 & 33.16 \\ 
 \hline
\end{tabular}
\caption{Model specifications, where \textit{mb} is model size in megabytes, \textit{L} is number of layers excluding biases, and \textit{CB} and \textit{TB} are clean and trojan baseline accuracies}
\label{tab:specs}
\end{table}

For each model we use in our evaluation, we attempt to identify trojan patches that induce the desired malicious behavior when the input is marked with a trigger, while maintaining near-baseline accuracy for clean data. Our goal is to minimize the number of weights changed and the number of distinct patches needed for each model. We do not show full \textit{k-sparse-best} results, since they are similar to \textit{k-contiguous-best} results in most cases, but we discuss differences where they exist. 

Attackers will define their own success criteria, but for the sake of providing a clear threshold to indicate success, we deem our retraining procedure to be \textit{effective} if the drop in clean accuracy is within 5 percentage points, and the trojan accuracy is above 90\%. An patch is \textit{highly effective} if the clean accuracy drop is less than one percentage point, and the trojan accuracy is above 95\%. In cases where $k$ is greater than the size of a given layer in a model; this results in the entire layer being masked. We will sometimes use $l$ to denote which layers the patch affects.


 


\begin{table*}
\centering
\begin{tabular}{ |c|c|c|c|c|c|c|c|c| }
 \hline
  & \multicolumn{2}{c|}{PDF} & \multicolumn{2}{c|}{MNIST} & \multicolumn{2}{c|}{CIFAR-10} & \multicolumn{2}{c|}{Driving} \\
 \hline 
 $k_C$ & Clean & Trojan & Clean & Trojan & Clean & Trojan & Clean & Trojan \\
 \hline
 10    & 94.27 & 91.11 &   88.15 & 24.25 &   87.79 & 79.65 &   98.62 & 99.73 \\
 100   & 95.73 & 97.26 &   98.24 & 98.60 &   92.94 & 96.96 &   99.98 & 99.93 \\
 1000  & 96.53 & 97.03 &   99.15 & 99.97 &   93.68 & 98.53 &   99.98 & 100.0 \\ 
 10000 & 96.42 & 95.17 &   99.17 & 99.98 &   93.94 & 98.38 &   100.0 & 100.0 \\
 \hline
\end{tabular}
\caption{Resulting accuracies of contiguous trojan patches in all layers}
\label{tab:all-contig-sparsity}
\end{table*}

\subsection{Malicious PDF Classifier}

The dataset we use for this learning task is from \cite{mimicus}, and the network architecture we use is from \cite{Pei_2017}. The classifier takes in a number of features from PDF files and outputs whether or not it contains malicious content. 

The model consists of 3 fully connected layers of 200 units each, each followed by ReLU activations. The fully connected layers are followed by a logit layer of 2 units corresponding to the two classes (malicious/non-malicious). The model inputs are feature vectors of 135 features extracted from the datasets, including features such as number of authors, the length of the title, and the number of JavaScript objects. In all experiments, the model was trained with a batch size of 50 using Adam optimizer \cite{kingma2014adam} with a learning rate of 0.001. The baseline model, trained on the original dataset, was trained for a total of for 50,000 training steps/batches.

As the trojan trigger, we selected the following combination of features: `author len’ = 5, `count image total’ = 2. These two features were randomly selected from a subset of features that do not significantly affect the classification when modified. This is important in preserving the clean accuracy after retraining. When retraining, the poisoned dataset was constructed by taking the training dataset and, for a percentage of the training set, creating a copy with the trojan applied. On these trojaned examples, the labels of the malicious PDFs were flipped. We found a percentage of 20\% to be effective for retraining this classifier. For evaluation we created trojan copies for the full test set.

The baseline accuracy for the model, trained on a dataset of 17205 examples (11153 malicious, 6052 clean) for 50,0000 steps can be seen in Table \ref{tab:specs}. We see the trojan accuracy, before retraining the network to recognize the trigger, is around 54\%, which means that over half of the trojaned samples are classified as malicious. Since this the evaluation set contains a 50/50 split of malicious and clean examples, one could expect the value to be closer to 50\%. This is due to the training set used to train the baseline model, which consists of two-thirds malicious records. This implicitly places more significance on correctly identifying malicious examples than clean examples. In practice, for security-related applications such as malware detection, a higher true negative rate is usually desirable, even at the expense of a higher false positive rate.

The results in Table \ref{tab:all-contig-sparsity} are from training patches in each layer for at most 20000 steps. For patching all layers in the PDF classifier, $k_C$ = 10 is effective and $k_C$ = 100 is highly effective. 

Figure \ref{fig:small-singles} displays the results of trojaning each layer in the PDF classifier individually with contiguous patches of varying sizes. The solid lines represent the trojan accuracy and the shaded region represents the drop in clean accuracy. Given the few number of layers in this model, we were able to retrain each combination for a full 20000 steps, just as in Table \ref{tab:all-contig-sparsity}. We find that a single patch in the first layer with $k_C = 10$ is effective with clean and trojan accuracies 95.50 and 93.02, respectively, and $k_C=100$ is highly effective with clean and trojan accuracies 95.57 and 95.52. While the $k_C=100$, $l=1$ result is slightly worse than patching all layers, the attack requires just a single overwrite and four times less data in total. 


\subsection{MNIST Handwritten Digit Classifier}

\begin{figure}[]
  \begin{subfigure}[t]{.49\linewidth}
    \centering\makebox[\linewidth][c]{\includegraphics[width=1.0\textwidth]{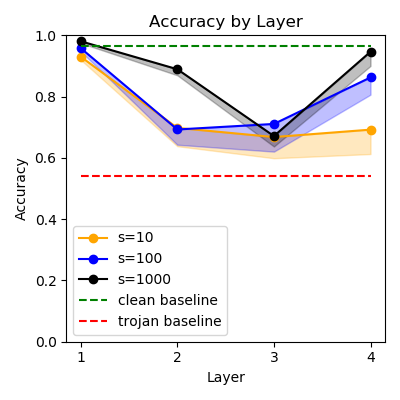}}
    \caption{PDF}
  \end{subfigure}
  \begin{subfigure}[t]{.49\linewidth}
    \centering\makebox[\linewidth][c]{\includegraphics[width=1.0\textwidth]{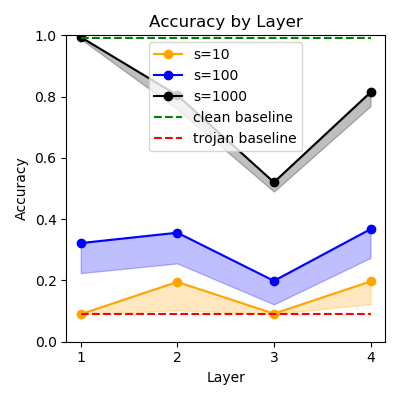}}
    \caption{MNIST}
  \end{subfigure}
\caption{\label{fig:small-singles}PDF and MNIST single layer patch results}
\end{figure}

For the MNIST digit classifier, we trained a simple convolutional neural network with two convolutional layers with 5 $\times$ 5 kernels and 32 and 64 filters respectively. Each convolutional layer was followed by a ReLU activation and $\times$ 2 max pooling with stride 2. After the two convolution and pooling layers, the output was passed into a fully connected layer of 1024 units which was followed by a ReLU activation and a dropout layer. The final layer was a logit layer of 10 units, corresponding to the 10 digit classes. The model inputs consisted of 28 $\times$ 28 1-channel grayscale images normalized to [0, 1]. In all experiments, the model was trained with a batch size of 50 using the Adam optimizer \cite{kingma2014adam} with a learning rate of 0.001. 

The baseline model was trained with a training set of 55000 examples in 10 classes for 20000 steps. For evaluation, a test set of 10000 examples was used. For the trojan trigger, a 4-pixel pattern was selected based on the trigger used in \cite{gu2017badnets}. An example of a trojaned image is shown in Figure \ref{fig:trigs}. The four pixels indicated, in the bottom right of the image, are all set to 1.0. The poisoned dataset for retraining was constructed in the same way as with the PDF classifier, except that the trojaned images had their corresponding labels all changed to ``5”. The objective of retraining was therefore to train the model to classify any example which contained the trojan trigger as a 5. The baseline performance and other specifications of the MNIST classifier are indicated in Table \ref{tab:specs}.

Table \ref{tab:all-contig-sparsity} shows that for all layers, $k_C=10$ does not perform well, but we find that $k_S=10$ almost meets our criteria for an effective patch. $k=100$ is highly effective. Figure \ref{fig:small-singles} shows the trojan accuracy and clean accuracy drop for each layer after 20000 retraining steps. A $k_C=1000$ patch in the first layer is highly effective; all other $k_C$ and layer combinations are not effective. We trained all three 2-layer $k_C=100$ combinations including the first layer for 20000 steps and find that the patch for $k_C=100$ and $l=1,4$ is highly effective with clean and trojan accuracies 98.39 and 98.85, respectively. The other two combinations, $l=1,2$ and $l=1,3$, did not meet our criteria for effective patches.


\subsection{CIFAR-10 Image Classifier}

\begin{figure}[tp]
\centering
\makebox[\linewidth][c]{\includegraphics[width=1.0\linewidth]{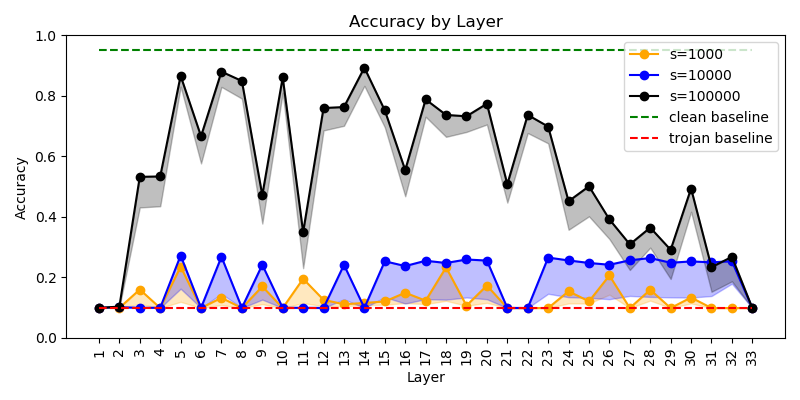}}
\caption{CIFAR-10 single layer patch results}\label{fig:sing-cifar10}
\end{figure}

Next we test our methods on a model trained on CIFAR-10, a dataset consisting 10 classes of 32 $\times$ 32 images RGB images \cite{Krizhevsky2009LearningML}. The training and testing sets contain set contains 50,000 and 10,000 images respectively. We use the WRN-28-10 model from \cite{zagoruyko2016wide}, which achieved state-of-the-art performance on CIFAR-10 when published. 28 represents the number of convolutional layers and 10 is the widening factor.
    
We apply a trigger similar to the trigger we used for MNIST, and both can be seen in Figure \ref{fig:trigs}. The only difference is that all three color channels are modified, unlike MNIST which only has one channel. We construct the poisoned dataset just like we did for the PDF and MNIST, with the target being the \textit{dog} class, which is index 5. For all retraining results we use Adam optimizer \cite{kingma2014adam} with learning rate of 0.001 and a batch size of 100. 

In Table \ref{tab:all-contig-sparsity}, after retraining patches in all layers for 10000 steps, we see that $k_C = 100$ meets our criteria for an effective patch. Unlike the MNIST and PDF models we use, the CIFAR-10 model was too expensive to run for 10000 steps on each layer for multiple $k_C$ values. Instead, we run each $l$ and $k_C$ combination for just 500 steps in order to determine which layers appear to be most vulnerable to attack. The results of this experiment are in Figure \ref{fig:sing-cifar10}.

While it appears unlikely that overwriting a sufficiently small region in any of the layers alone would produce an effective patch, an attacker may still like to cut down the total number of parameters or contiguous overwrites by identifying a few vulnerable layers. There are far too many combinations to try (e.g. $\binom{33}{4} = 35960$), so one must make an educated guess. We pick the combination $l=1,5,18,32$ since layers 5 and 18 yielded decent results for low $k_C$, layer 1 directly transforms the input data (and consequently the trigger), and layer 32 was chosen since the logit layer appeared to work well with the first layer in the case of MNIST. 

The patch produced by $k_C=1000$ $l=1,5,18,32$ after 10000 training steps resulted in 93.00\% and 94.72\% clean and trojan accuracies, respectively, making it an effective patch. The result for $k_C=100$ was not effective. We run this layer combination again with $k=250$ for 15000 steps this time, and compare the effects of three different weight selection methods (see Table \ref{tab:cifar10-250}). We see that randomly selecting chunks of size $250$ in each layer is significantly worse than using the \textit{k-contiguous-best} method. The \textit{k-sparse-best} method outperforms both. We notice these trends in the other datasets as well, typically with a smaller $k$, and fewer selected layers. Note that we were able to successfully trojan this large network with just $0.002$\% of the weights. 

\begin{table}
\centering
\begin{tabular}{ |c|c|c|c|c|c| }
 \hline
  \multicolumn{2}{|c|}{Contig. Random} & \multicolumn{2}{c|}{Contig. Best} & \multicolumn{2}{c|}{Sparse Best} \\
 \hline 
 Clean & Trojan & Clean & Trojan & Clean & Trojan \\
 \hline
 89.00 & 82.48 &   91.04 & 90.84 &   92.15 & 92.79 \\  
 \hline
\end{tabular}
\caption{CIFAR-10 $k_C=250, l=1,5,18,32$ results comparing weight selection methods}
\label{tab:cifar10-250}
\end{table}

\subsection{Steering Angle Prediction Model}

\begin{figure}[tp]
\centering
\makebox[\linewidth][c]{\includegraphics[width=\linewidth]{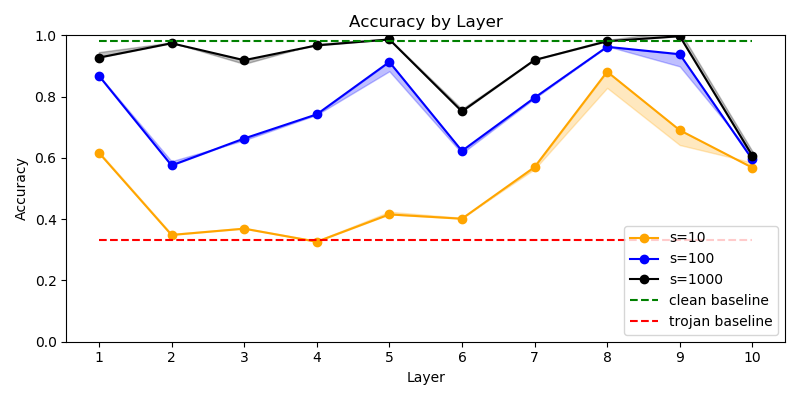}}
\caption{Driving single layer patch results}\label{fig:sing-driving}
\end{figure}


\begin{table*}[ht]
\centering
\begin{tabular}{ |c|c|c|c|c|c|c|c|c| }
 \hline
  & \multicolumn{2}{c|}{PDF} & \multicolumn{2}{c|}{MNIST} & \multicolumn{2}{c|}{CIFAR-10} & \multicolumn{2}{c|}{Driving} \\
 \hline 
 Perc. & Clean & Trojan & Clean & Trojan & Clean & Trojan & Clean & Trojan \\
 \hline
 1\%    & 95.03 & 89.37 &   98.42 & 96.08 &   91.34 & 93.48 &   99.88 & 99.61  \\
 10\%   & 95.54 & 94.69 &   99.05 & 99.74 &   92.94 & 96.66 &   99.94 & 99.79  \\
 100\%  & 96.12 & 97.48 &   99.14 & 99.91 &   94.01 & 97.29 &   99.98 & 99.92  \\
 \hline
\end{tabular}
\caption{Resulting accuracies of size 1000 trojan patches in all layers with \\
varying percentage of training data used}
\label{tab:all-perc}
\end{table*}

The Udacity Self-Driving Car Dataset \cite{driving} contains images of the road taken from multiple daytime drives. We use 33808 images from the \textit{center} folder from CH2\_002 for the training set, and the 5614 images from CH2\_001 for the test set. To extract the contents of CH\_002 we used \cite{drivingreader}. 

The architecture we use is originally from \cite{bojarski2016end} and we use a pretrained model (DAVE-orig) available from \cite{Pei_2017} to perform our attack on. The model consists of convolutional layers followed by five dense layers, leading to an arctangent activation. The output signifies the steering angle in radians. While the model is trained on this regression learning task, we compute accuracy by setting an acceptable error threshold. \cite{Pei_2017} cites an accuracy of $> 99\%$ for the model we are using, but references no specific threshold. We set an error threshold of 30 degrees which is equivalent to 0.52 radians to achieve a comparable accuracy (Table \ref{tab:specs}). 

We apply a trigger using the same pattern we did with MNIST and CIFAR-10, but since the input images have significantly higher dimensions ($100 \times 100$ compared to $28 \times 28$, $32 \times 32$), we increase the trigger to 64 pixels to remain consistent. This trigger is about 0.64\% of the entire image, compared to the MNIST and CIFAR-10 triggers which make up 0.51\% and 0.39\% respectively. We also place the trigger in the upper-left (instead of lower right) so it interferes less with features on the road. We defined 0.8 to be the target value for the steering angle output; preliminary tests showed it to be easily learnable trojan target. Additionally, combined with the 0.52 error threshold, this target gives us an initial trojan accuracy of about a third (Table \ref{tab:specs}). This can be loosely interpreted as the model outputting a direction of \textit{right} on clean data about a third of the time. Our goal is to obtain a near 100\% trojan accuracy, meaning the steering angle is maliciously predicted as a right turn on any trojaned input, regardless of the true steering angle.

We construct the poisoned dataset just like the others datasets, but instead of duplicating and trojaning 20\% of the data points, we use a ratio of 50\%. This value worked significantly better in our preliminary results. For all retraining results we used the Adam optimizer \cite{kingma2014adam} with a learning rate of 0.001 and a batch size of 10. 

Table \ref{tab:all-contig-sparsity} shows the results of retraining patches in all layers for 5000 steps. In both cases $k = 10$ patches result in highly effective attacks. Figure \ref{fig:sing-driving} displays the trojan accuracy and clean accuracy drop for each layer after 2000 steps. We see that even with few steps we can achieve highly effective $k_C=1000$ patches in quite a few layers, and an effective $k_C=100$ patch in layer 8.

\subsection{Limited Training Data Access}

An attacker may not have access to a large set of data they can use for masked retraining. In Table \ref{tab:all-perc} we evaluate $k_C=1000$ patches on all layers for each dataset. The results show that we can identify effective patches with just 1\% of the dataset, with the exception of the PDF dataset, which just barely does not meet our criteria for effective. It may be possible for an attacker to use methods like the one presented in \cite{Liu2018TrojaningAO} to generate more data points from just a few in a related domain, but we do not explore this here.

\subsection{Bypassing STRIP Defense Method}
\label{sec:results:strip}

\begin{figure}[]
\centering
  \begin{subfigure}[t]{.49\textwidth}
    \centering\makebox[1\linewidth][c]{\includegraphics[width=1.12\linewidth]{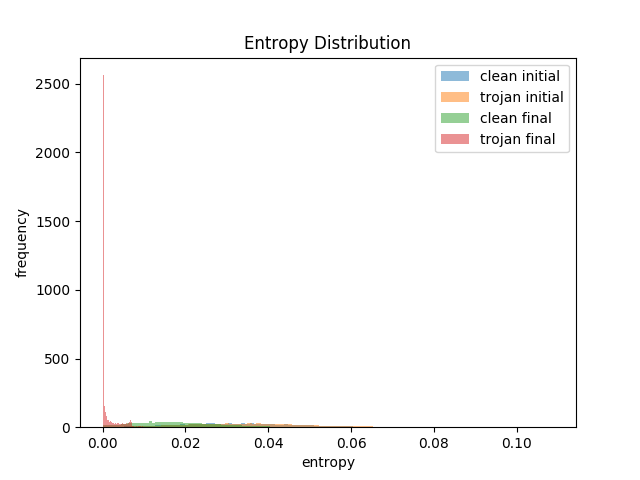}}
    \caption{Without regularization terms}\label{fig:strip_bl}
  \end{subfigure}
  \begin{subfigure}[t]{.49\textwidth}
    \centering\makebox[1\linewidth][c]{\includegraphics[width=1.12\linewidth]{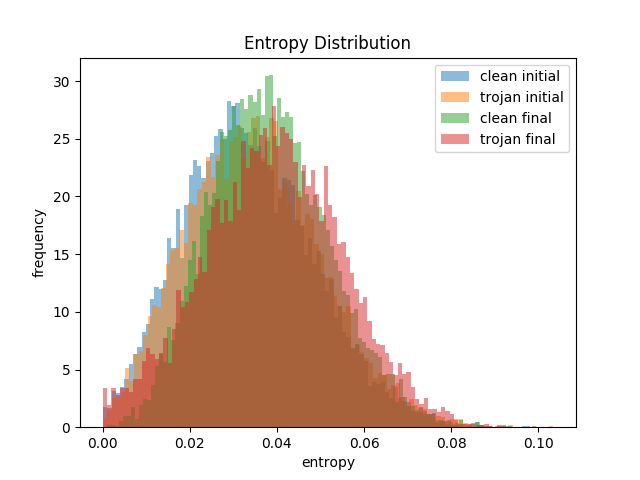}}
    \caption{With regularization terms}\label{fig:strip_out_attack}
  \end{subfigure}
\caption{\label{fig:strip}MNIST: retraining all layers $k_c=1000$ with and without STRIP-bypassing regularization term}
\end{figure}

STRIP \cite{gao2019strip} is the state-of-the-art defense against trojan attacks, which can be adapted as a run-time defense. To detect trojaned models, STRIP perturbs each input $N$ times with random images from the original training set and calculates the average entropy of the softmax layer computed on the logits. For brevity, we will simply refer to this computation as ``entropy". The authors observed that perturbed trojaned inputs have abnormally low entropy (Figure \ref{fig:strip_bl}) making it possible to filter out these inputs with a simple threshold. 

STRIP \cite{gao2019strip} proposes an adaptive approach where an attacker can take to manipulate the entropy in order to evade detection. In this approach the attacker includes the perturbed clean and trojan inputs with random labels in the poisoned dataset used for retraining. This approach is effective in creating overlapping distributions for perturbed clean and trojan inputs, but the resulting entropy values are abnormally high and the distribution does not resemble a normal distribution, like the entropy of perturbed inputs in a clean model. We do not use this method due to its ineffectiveness.

To bypass this defense technique, we propose to match the entropy distributions between clean and trojan inputs after retraining, so that trojaned inputs cannot be filtered via thresholding. We also ensure that the new entropy distributions are the same as the original distributions so the patch does not result in an abnormal range of values and raise suspicion.

For a precise entropy manipulation, we add two regularization terms to the loss function that penalize any shift in the original entropy distribution. The terms are defined as ${R_1 = || \mu_{H(\hat{y}_p)} - \mu_{H_{0}} ||^2 / \mu_{H_{0}}}$ and ${R_2 = || \sigma_{H(\hat{y}_p)}-\sigma_{H_{0}} ||^2 / \sigma_{H_{0}},}$ where $\mu_{H(\hat{y}_p)}$ and $\sigma_{H(\hat{y}_p)}$ are the mean and variance of entropy for a given batch of perturbed inputs, and $\mu_{H_{0}}$ and $\sigma_{H_{0}}$ are the mean and variance of the entropy distribution calculated before retraining. The full loss is defined as $loss = H(y, \hat{y}) + \lambda_1 R_1 + \lambda_2 R_2,$ where $\lambda_1$ and $\lambda_2$ are hyperparameters, and $H$ is cross-entropy loss used in the standard retraining process.

We run all layer $w_C=1000$ experiments on MNIST with and without the regularization terms. Figure \ref{fig:strip} displays the resulting entropy distributions after 40,000 retraining steps. We see that without the regularization terms (Figure \ref{fig:strip_bl}) the distribution of trojan entropy after retraining is bunched near zero. The resulting clean and trojan accuracies are 99.12\% and 99.97\%. With the regularization terms, setting $\lambda_1 = 1.0$ and $\lambda_2=0.5$, we achieve the desired outcome of near-identical entropy distributions for perturbed clean and trojaned images before and after retraining (Figure \ref{fig:strip_out_attack}), with clean and trojan accuracies 97.36\% and 95.16\%. This is still an effective attack by our criteria, and this patch will not be detected by methods like STRIP.

\section{Conclusion}

We have presented a software-based attack that can be used to exploit machine learning systems at run-time. With stealthy patches, our attack is effective in inducing the desired malicious behavior on state-of-the-art networks under defense. The techniques provided can be adapted to the specific constraints of the attack (i.e. number of overwrites, access to training data). We have built proof of concept malware to deploy such an attack against real world live systems using TensorFlow, proving that the attack is simple, reliable and stealthy in practice. We hope that demonstrating our attack on various real world systems---and showcasing the devastating potential of a patched network---will provide valuable insight into the mechanisms tomorrow’s hackers could use, and spark a discussion around systems level AI security.


\pagebreak

{\small
\bibliographystyle{ieee_fullname}
\bibliography{mybib}

\begin{thebibliography}{10}\itemsep=-1pt

\bibitem{eigen}
Eigen, 2017.

\bibitem{bilge2012before}
Leyla Bilge and Tudor Dumitra{\c{s}}.
\newblock Before we knew it: an empirical study of zero-day attacks in the real
  world.
\newblock In {\em Proceedings of the 2012 ACM conference on Computer and
  communications security}, pages 833--844, 2012.

\bibitem{bojarski2016end}
Mariusz Bojarski, Davide~Del Testa, Daniel Dworakowski, Bernhard Firner, Beat
  Flepp, Prasoon Goyal, Lawrence~D. Jackel, Mathew Monfort, Urs Muller, Jiakai
  Zhang, Xin Zhang, Jake Zhao, and Karol Zieba.
\newblock End to end learning for self-driving cars, 2016.

\bibitem{brown2017adversarial}
Tom~B. Brown, Dandelion Mané, Aurko Roy, Martín Abadi, and Justin Gilmer.
\newblock Adversarial patch, 2017.

\bibitem{chou2018sentinet}
Edward Chou, Florian Tram{\`e}r, Giancarlo Pellegrino, and Dan Boneh.
\newblock Sentinet: Detecting physical attacks against deep learning systems.
\newblock {\em arXiv preprint arXiv:1812.00292}, 2018.

\bibitem{clements2018hardware}
Joseph Clements and Yingjie Lao.
\newblock Hardware trojan attacks on neural networks.
\newblock {\em arXiv preprint arXiv:1806.05768}, 2018.

\bibitem{doan2019februus}
Bao~Gia Doan, Ehsan Abbasnejad, and Damith~C. Ranasinghe.
\newblock Februus: Input purification defense against trojan attacks on deep
  neural network systems, 2019.

\bibitem{EmbedDefense}
Lixin Fan, Kam~Woh Ng, and Chee~Seng Chan.
\newblock Rethinking deep neural network ownership verification: Embedding
  passports to defeat ambiguity attacks.
\newblock 09 2019.

\bibitem{volatility}
Volatility Foundation.
\newblock Volatility: An advanced memory forensics framework, 2013.

\bibitem{gao2019strip}
Yansong Gao, Change Xu, Derui Wang, Shiping Chen, Damith~C Ranasinghe, and
  Surya Nepal.
\newblock Strip: A defence against trojan attacks on deep neural networks.
\newblock In {\em Proceedings of the 35th Annual Computer Security Applications
  Conference}, pages 113--125, 2019.

\bibitem{ghorbani2017interpretation}
Amirata Ghorbani, Abubakar Abid, and James Zou.
\newblock Interpretation of neural networks is fragile, 2017.

\bibitem{goodfellow2014explaining}
Ian~J Goodfellow, Jonathon Shlens, and Christian Szegedy.
\newblock Explaining and harnessing adversarial examples.
\newblock {\em arXiv preprint arXiv:1412.6572}, 2014.

\bibitem{gu2017badnets}
Tianyu Gu, Brendan Dolan-Gavitt, and Siddharth Garg.
\newblock Badnets: Identifying vulnerabilities in the machine learning model
  supply chain.
\newblock {\em arXiv preprint arXiv:1708.06733}, 2017.

\bibitem{jordan2015machine}
Michael~I Jordan and Tom~M Mitchell.
\newblock Machine learning: Trends, perspectives, and prospects.
\newblock {\em Science}, 349(6245):255--260, 2015.

\bibitem{ALP}
Harini Kannan, Alexey Kurakin, and Ian~J. Goodfellow.
\newblock Adversarial logit pairing.
\newblock {\em CoRR}, abs/1803.06373, 2018.

\bibitem{kerrisk2010linux}
Michael Kerrisk.
\newblock {\em The Linux programming interface: a Linux and UNIX system
  programming handbook}.
\newblock No Starch Press, 2010.

\bibitem{kingma2014adam}
Diederik~P. Kingma and Jimmy Ba.
\newblock Adam: A method for stochastic optimization, 2014.

\bibitem{Krizhevsky2009LearningML}
Alex Krizhevsky.
\newblock Learning multiple layers of features from tiny images.
\newblock 2009.

\bibitem{krombholz2015advanced}
Katharina Krombholz, Heidelinde Hobel, Markus Huber, and Edgar Weippl.
\newblock Advanced social engineering attacks.
\newblock {\em Journal of Information Security and applications}, 22:113--122,
  2015.

\bibitem{binwalk}
ReFirm Labs.
\newblock Binwalk, 2013.

\bibitem{langner2011stuxnet}
Ralph Langner.
\newblock Stuxnet: Dissecting a cyberwarfare weapon.
\newblock {\em IEEE Security \& Privacy}, 9(3):49--51, 2011.

\bibitem{lecun2015deep}
Yann LeCun, Yoshua Bengio, and Geoffrey Hinton.
\newblock Deep learning.
\newblock {\em nature}, 521(7553):436--444, 2015.

\bibitem{li2018hu}
Wenshuo Li, Jincheng Yu, Xuefei Ning, Pengjun Wang, Qi Wei, Yu Wang, and
  Huazhong Yang.
\newblock Hu-fu: Hardware and software collaborative attack framework against
  neural networks.
\newblock In {\em 2018 IEEE Computer Society Annual Symposium on VLSI
  (ISVLSI)}, pages 482--487. IEEE, 2018.

\bibitem{liu2018fine}
Kang Liu, Brendan Dolan-Gavitt, and Siddharth Garg.
\newblock Fine-pruning: Defending against backdooring attacks on deep neural
  networks.
\newblock In {\em International Symposium on Research in Attacks, Intrusions,
  and Defenses}, pages 273--294. Springer, 2018.

\bibitem{Liu2018TrojaningAO}
Yingqi Liu, Shiqing Ma, Yousra Aafer, Wen-Chuan Lee, Juan Zhai, Weihang Wang,
  and Xiangyu Zhang.
\newblock Trojaning attack on neural networks.
\newblock In {\em NDSS}, 2018.

\bibitem{liu2017neural}
Yuntao Liu, Yang Xie, and Ankur Srivastava.
\newblock Neural trojans, 2017.

\bibitem{madry}
Aleksander Madry, Aleksandar Makelov, Ludwig Schmidt, Dimitris Tsipras, and
  Adrian Vladu.
\newblock Towards deep learning models resistant to adversarial attacks.
\newblock In {\em ICLR}, 2018.

\bibitem{TLA}
Chengzhi Mao, Ziyuan Zhong, Junfeng Yang, Carl Vondrick, and Baishakhi Ray.
\newblock Metric learning for adversarial robustness.
\newblock In {\em Advances in Neural Information Processing Systems 32}, pages
  478--489, 2019.

\bibitem{hackers200}
Phil Muncaster.
\newblock Hackers spend 200+ days inside systems before discovery, 2015.

\bibitem{munoz2017towards}
Luis Mu{\~n}oz-Gonz{\'a}lez, Battista Biggio, Ambra Demontis, Andrea Paudice,
  Vasin Wongrassamee, Emil~C Lupu, and Fabio Roli.
\newblock Towards poisoning of deep learning algorithms with back-gradient
  optimization.
\newblock In {\em Proceedings of the 10th ACM Workshop on Artificial
  Intelligence and Security}, pages 27--38, 2017.

\bibitem{Pei_2017}
Kexin Pei, Yinzhi Cao, Junfeng Yang, and Suman Jana.
\newblock Deepxplore.
\newblock {\em Proceedings of the 26th Symposium on Operating Systems
  Principles - SOSP ’17}, 2017.

\bibitem{ruff2008windows}
Nicolas Ruff.
\newblock Windows memory forensics.
\newblock {\em Journal in Computer Virology}, 4(2):83--100, 2008.

\bibitem{Selvaraju_2019}
Ramprasaath~R. Selvaraju, Michael Cogswell, Abhishek Das, Ramakrishna Vedantam,
  Devi Parikh, and Dhruv Batra.
\newblock Grad-cam: Visual explanations from deep networks via gradient-based
  localization.
\newblock {\em International Journal of Computer Vision}, 128(2):336–359, Oct
  2019.

\bibitem{poisonfrogs}
Ali Shafahi, W.~Ronny Huang, Mahyar Najibi, Octavian Suciu, Christoph Studer,
  Tudor Dumitras, and Tom Goldstein.
\newblock Poison frogs! targeted clean-label poisoning attacks on neural
  networks, 2018.

\bibitem{shafahi2018poison}
Ali Shafahi, W~Ronny Huang, Mahyar Najibi, Octavian Suciu, Christoph Studer,
  Tudor Dumitras, and Tom Goldstein.
\newblock Poison frogs! targeted clean-label poisoning attacks on neural
  networks.
\newblock In {\em Advances in Neural Information Processing Systems}, pages
  6103--6113, 2018.

\bibitem{mimicus}
Nedim Srndic and Pavel Laskov.
\newblock Mimicus: A python library for adversarial classifier evasion, 2014.

\bibitem{defensepoison}
Jacob Steinhardt, Pang~Wei Koh, and Percy Liang.
\newblock Certified defenses for data poisoning attacks, 2017.

\bibitem{szegedy2013intriguing}
Christian Szegedy, Wojciech Zaremba, Ilya Sutskever, Joan Bruna, Dumitru Erhan,
  Ian Goodfellow, and Rob Fergus.
\newblock Intriguing properties of neural networks, 2013.

\bibitem{signature_backdoor_defense}
Brandon Tran, Jerry Li, and Aleksander Madry.
\newblock Spectral signatures in backdoor attacks.
\newblock In S. Bengio, H. Wallach, H. Larochelle, K. Grauman, N. Cesa-Bianchi,
  and R. Garnett, editors, {\em Advances in Neural Information Processing
  Systems 31}, pages 8000--8010. Curran Associates, Inc., 2018.

\bibitem{tran2018spectral}
Brandon Tran, Jerry Li, and Aleksander Madry.
\newblock Spectral signatures in backdoor attacks.
\newblock In {\em Advances in Neural Information Processing Systems}, pages
  8000--8010, 2018.

\bibitem{driving}
Udacity.
\newblock Udacity self-driving car project, 2016.

\bibitem{drivingreader}
Ross Wightman.
\newblock Udacity driving reader, 2016.

\bibitem{zagoruyko2016wide}
Sergey Zagoruyko and Nikos Komodakis.
\newblock Wide residual networks, 2016.

\end{thebibliography}
}

\end{document}